\documentclass[a4paper, 12pt]{article}
\usepackage{amssymb,amsfonts,amsmath,mathtext,enumerate,float}

  \renewenvironment{thebibliography}[1]{%
    \begin{oldthebibliography}{#1}%
      \setlength{\parskip}{0ex}%
      \setlength{\itemsep}{0ex}%
  }%
  {%
    \end{oldthebibliography}%
  }
\usepackage{epsfig}
\begin{document}
\begin{center}
\begin{Large}
%\noindent
{\bf{DISIM${_{b}}$(2) Local Relativistic Symmetry\\ and\\[2mm] 
 Finslerian Extension of the Theory of Relativity}}
\end{Large}
\vskip 7mm
%\noindent
{\bf{George Yu. Bogoslovsky}}
\vskip 3mm
%\hfill \begin{minipage}{15.95cm}
\begin{small}
%\noindent
Skobeltsyn Institute of Nuclear Physics, Lomonosov Moscow State University, Moscow, Russia\\
bogoslov@theory.sinp.msu.ru\\
\end{small}
\end{center}
\vskip 3mm
\begin{small}
\noindent
{\bf{Abstract.}}\,\, Finslerian extension of the theory of relativity implies that space-time can be not only  in an amorphous state which is described by Riemann geometry
but also in ordered, i.e. crystalline states which are described by Finsler geometry. Transitions between
various metric states of space-time have the meaning of phase transitions in its geometric structure. These
transitions together with the evolution of each of the possible metric states make up the general picture of space-time
manifold dynamics. It is shown that there are only two types of curved Finslerian spaces endowed with local
relativistic symmetry. However the metric of only one of them satisfies the correspondence principle with
Riemannian metric of the general theory of relativity and therefore underlies viable Finslerian extension of the GR.
Since the existing purely geometric approaches to a Finslerian generalization of Einstein's equations do not allow one
to obtain such generalized equations which would provide a local relativistic symmetry of their solutions,
special attention is paid to the property of the specific invariance of viable Finslerian metric under local conformal
transformations of those fields on which it explicitly depends. It is this property that makes it possible to use
the well-known methods of conventional field theory and thereby to circumvent the above-mentioned difficulties
arising within the framework of purely geometric approaches to a Finslerian generalization of  Einstein's equations.
\\
 %\vskip 1mm
 
%\\
\end{small}
\vskip 5mm
\noindent
%{\large{\bf 1\,. Introduction}}
{\bf 1.\,Introduction}\\
%\vskip 1mm

\hspace{3mm}The impressive successes of the contemporary cosmology are based  to a large extent on the use of the Riemannian space-time model and the Einstein equations
with $\Lambda$ term
\begin{equation}\label{1}
R_{ik}-\frac{1}{2}Rg_{ik}=\Lambda g_{ik}+T_{ik}\,.
\end{equation}
In connection with the $\Lambda$ term let us direct our attention to the Weyl theory [1, 2]. Trying to construct a unified theory of gravitation and electromagnetism, Weyl proceeded from the requirement of invariance of the theory with respect to the local conformal transformations of the Riemannian metric tensor $g_{ik}\ \to\exp [2\sigma (x)]\,g_{ik}\,. $
The appearing Abelian gauge field, denoted below by $\,A_i\,$, was identified by Weyl with an electromagnetic field. The basis for such identification was the gradient character of the corresponding gauge transformations
$A_i\,\to\,A_i+a{\sigma}_{;i}\,,$ where $\,a\,$ is a constant with the dimension of length.
It is easy, however, to show [3] that in reality in the Weyl theory the $\,A_i\,$  is not an electromagnetic field but acquires mass due to the violation of local conformal invariance.
This result is essentially contained in the Weyl variational principle
\begin{equation}\label{2}
\delta\left [\frac{1}{16\Lambda} \int{\!\!\left (R+\frac{6}{a}A^i_{;i}-\frac{6}{a^2}A_iA^i\right )}^{\!\!2}\!\!\sqrt{-g}\,d^{\,4}x-\frac{1}{4}\int\!\!F_{ik}F^{ik}\sqrt{-g}\,d^{\,4}x\right ]\!\!=0\,,
\end{equation}
where $F_{ik}=A_{k;i}-A_{i;k}$ and $\Lambda $ is a constant.
In fact, using the Weyl gauge condition
\begin{equation}\label{3}
R+\frac{6}{a}A^i_{;i}-\frac{6}{a^2}A_iA^i=-4\Lambda
\end{equation}
and his variational principle (2), we get the following system of equations
\begin{equation}\label{4}
{F^{ik}}_{;k}-\frac{6}{a^2}A^i=0\,,
\end{equation}
\begin{equation}\label{5}
R_{ik}-\frac{1}{2}Rg_{ik}-\Lambda g_{ik}=-F_{il}{F^l}_k+\frac{1}{4}F_{lm}F^{lm}g_{ik}+\frac{6}{a^2}(A_iA_k-\frac{1}{2}A_lA^lg_{ik})\,.
\end{equation}
Since, by virtue of (4), the equality  $\,A^i_{;i}= 0\,$  holds and, by virtue of (5), the equality $\,R-\frac{6}{a^2}A_iA^i=-4\Lambda\,$  holds,
 it is clear that the Weyl gauge condition (3)
is fulfilled for the solutions of the system of equations (4) and (5).
Thus, a system of fourth-order field equations, following from the Weyl gauge invariant variational principle (2), together with the condition (3) fixing the gauge, is equivalent to the system of equations (4) and (5). In other words, the fixing of the gauge, which implies in the Weyl theory the incorporation of  length and mass standards, converts the Abelian vector field $A_i$  into a massive vector field and reduces the equations for gravitational field to the Einstein equations with $\Lambda$ term.

Let us return to the Einstein equations (1).
The vacuum Lorentz-invariant energy-momentum tensor $\,\Lambda g_{ik}\,$  can be written in the form
$\,\Lambda g_{ik}= diag ( {\rho}_v,\, -{\rho}_v,\, -{\rho}_v,\, -{\rho}_v )\,,$
where the so-called dark energy density  ${\rho}_v$  is responsible for the accelerated expansion of the Universe.
Thus, the corresponding vacuum space turns out to be filled with specific medium (\,scalar condensate, or quintessence\,), the pressure in which is negative and determined by the equation of state $p=-{\rho}_v$\,.

Apart from scalar condensate, the dynamic rearrangement of vacuum of quantized fields under spontaneous violation of initial gauge symmetry may give rise to anisotropic condensates.
The flat space-time keeps being the Minkowski space in the only case of scalar condensate. But, if an anisotropic condensate arises, the corresponding anisotropy also appears in space-time [4]. Speaking of local anisotropy of space-time we have in mind its Finslerian geometric structure.
In contrast to Riemann metrics, Finsler ones do not reduce to a quadratic form in the differentials of the coordinates.  Being homogeneous functions of the differentials of the coordinates of the second degree of homogeneity, such metrics serve as a natural generalization of Riemann metrics.

In order to arrive naturally at the flat relativistically invariant Finslerian space-time we first confine ourselves to a two-dimensional
space and show that it is possible to generalize the Lorentz transformations
\begin{equation}\label{6}
\left\{
\begin{array}{lll}
{x_0^{\prime}}&\!\!\!=&\!\!\!x_0\cosh \alpha -x\sinh \alpha \\
x'            &\!\!\!=-&\!\!\!x_0\sinh \alpha +x\cosh \alpha \,;
                       \qquad \tanh \alpha =v/c\\
\end{array}
\right.
\end{equation}
so that the new linear transformations will also form a group with
a single parameter $\,\alpha\,$ and will keep invariance of the wave
equation.
Guided by the conformal invariance of the electrodynamic
equations, we insert the additional scale transformations  $e^{-b\alpha}$   into the standard Lorentz ones (6). As a result we obtain the generalized Lorentz transformations in the form
\begin{equation}\label{7}
\left\{
\begin{array}{lclll}
   x'_0&\!\!\!=&\!\!\!e^{-b\alpha}&\!\!\!(&\!\!\!x_0\cosh\alpha -x\sinh\alpha)\\
   x'&\!\!\!=&\!\!\!e^{-b\alpha}&\!\!\!(-&\!\!\!x_0\sinh\alpha +x\cosh\alpha)\,,
\end{array}
\right.
\end{equation}
where $b$ is a dimensionless parameter of the scale transformations. Since
 the relation of the group parameter $\alpha$ to the velocity
$v$  of the primed frame remains unchanged, i.e. $\,\tanh\alpha =v/c\,,$ the generalized Lorentz transformations
 can be rewritten as follows

\smallskip
\begin{equation}\nonumber
\begin{large}
\left\{
\begin{array}{lcl}
x'_0&\!\!\!=&\!\!\!\left(\frac{1-v/c}{1+v/c}\right)^{\!b/2}
\frac{x_0-(v/c)x}{\sqrt{1-v^2/c^2}}\\
x'&\!\!\!=&\!\!\!\left(\frac{1-v/c}{1+v/c}\right)^{\!b/2}
\frac{x-(v/c)x_0}{\sqrt{1-v^2/c^2}}\,.
\end{array}
\right.
\end{large}
\end{equation}

\smallskip

\noindent
Obviously, in contrast to the standard Lorentz
transformations (6), the generalized ones (7) do not leave invariant the Minkowski metric $\,ds^2=dx_0^2-dx^2\,$
but conformally modify it. Therefore, the
question arises as to what the metric of an event space invariant
under such generalized Lorentz transformations is. A rigorous solution of this problem leads to the following metric

\smallskip
\begin{equation}\label{8}
ds^2=\left[\frac{(dx_0-dx)^2}{dx_0^2-dx^2}\right]^b(dx_0^2-dx^2)\,.
\end{equation}

\smallskip
\noindent
Not being a quadratic form but a homogeneous function of the coordinate
differentials of degree two, this metric  falls into the category of
Finslerian metrics. It describes a flat but anisotropic event
space.

    The observation made serves as a starting point for a viable Finslerian extension of both Special and General Relativity.
To this end, we should first generalize two-dimensional metric (8) to the case of a four-dimensional Finslerian event space.
As it turned out, there are two independent approaches to such a generalization. The first approach leads to a flat Finslerian
event space with partially broken 3D isotropy. The corresponding Finslerian metric, within the framework of the correspondence
principle, generalizes the pseudo-Euclidean metric of the Minkowski space and describes another possible vacuum state of a flat
space-time, namely, its axially symmetric crystalline state. Accordingly, the axially symmetric event space  underlies the viable
Finslerian extension of Special Relativity. The second approach  leads to a flat Finslerian event space with entirely broken 3D
isotropy. The corresponding Finslerian metric describes one more possible vacuum state of a flat space-time, namely, its entirely
anisotropic crystalline state. Such a state is considered in detail in Section 4. However, the main purpose of this article is to
describe the physically meaningful Finslerian extension of General  Relativity (see Section 5). At the same time, there is no need
to say that the corresponding Finslerian extension of Special Relativity should be considered first. Therefore, the following
Sections 2 and 3 are devoted specifically to the Finslerian Special Relativity. Despite a preliminary publication [5], these Sections are presented here in their original form.

Although this article has a review character, it contains (with few exceptions)
only those results on Finslerian extensions of the relativity theory, that were obtained by the
author or with his participation. As for the results obtained in the framework of the traditional,
purely geometric approach to a Finslerian modification of General Relativity, they are considered
in a recent review [6]. \\
%[6 L\"ammerzahl C.; Perlick V. Finsler geometry as a model for relativistic gravity.
%{\it Int. J. Geom. Methods Mod. Phys.} {\bf 2018}, 15, 1850166.]

\noindent
{\bf 2.\, Metric and group of relativistic symmetry of the flat Finslerian space-time
 %\phantom{wit}
with partially broken 3D isotropy}\\

Proceeding from the Finslerian metric of the two-dimensional space of events (8) and using the substitutions
$\,(dx_0^2-dx^2)\rightarrow (dx_0^2-d\boldsymbol x^{\,2})\,;\quad
(dx_0-dx)\rightarrow (dx_0-\boldsymbol n d\boldsymbol x)\,,
$ we arrive at the corresponding Finslerian metric [7] of the four-dimensional space of events
\begin{equation}\label{9}
ds^2=\left[\frac{(dx_0-\boldsymbol n d\boldsymbol
x)^2}{dx_0^2-d\boldsymbol x^{2}}\right]^b (dx_0^2-d\boldsymbol
x^{2}).
\end{equation}
Here the unit vector $\,\boldsymbol n\,$ indicates a preferred direction in 3D space while the parameter $\,b\,$ determines the magnitude of space anisotropy and characterizes the degree of deviation of the metric (9) from the Minkowski metric. Thus, the flat Finslerian  space-time (9) proves to be a generalization of the isotropic Minkowski space of conventional special relativity. Instead of the 3-parameter rotation group, the flat anisotropic event space (9) admits only 1-parameter group of rotations around the unit vector $\,\boldsymbol n\,$.  This is why one can speak of a partially broken 3D isotropy. As to relativistic
symmetry, it is realized by means of 3-parameter group of the generalized Lorentz transformations (\,of the generalized Lorentz boosts\,) which link the physically equivalent inertial reference frames in the flat anisotropic space-time  and leave its metric (9) invariant. The generalized Lorentz boosts have the form
\begin{equation}\label{10}
x'^i=D(\boldsymbol v,\boldsymbol n )R^i_j(\boldsymbol
v,\boldsymbol n )L^j_k(\boldsymbol v) x^k\,,
\end{equation}
where $\boldsymbol v$ denotes the velocities of moving (primed)
inertial reference frames, the matrices $L^j_k(\boldsymbol v)$
represent the ordinary Lorentz boosts, the matrices
$R^i_j(\boldsymbol v,\boldsymbol n )$ represent additional
rotations of the spatial axes of the moving frames around the
vectors $[\boldsymbol v\boldsymbol n ]$  through the angles
$$
\varphi=\arccos\left\{ 1-\frac{(1-\sqrt{1-\boldsymbol v^{2}/c^2})
[\boldsymbol v\boldsymbol n ]^2}{ (1-\boldsymbol v\boldsymbol n
/c)\boldsymbol v^{2}}\right\}
$$
of relativistic aberration of $\boldsymbol n,$ and the diagonal matrices
$$
D(\boldsymbol v,\boldsymbol n )=\left(\frac{1-\boldsymbol
v\boldsymbol n /c} {\sqrt{1-\boldsymbol v^{2}/c^2}} \right)^{\!\!b}I
$$
stand for the additional dilatational transformations of the event coordinates.

It should be noted that at $\,b=0\,$  the Finslerian metric (9) reduces to the Minkowski
metric, but the generalized Lorentz boosts, i.e. transformations (10) do not reduce to
ordinary Lorentz boosts and take the form
\begin{equation}\label{11}
x'^i=R^i_j(\boldsymbol
v,\boldsymbol n )L^j_k(\boldsymbol v) x^k\,.
\end{equation}
At $\,b=0\,,$ i.e. in the case of Minkowski space where all directions in 3D space are equivalent,
 $\,\boldsymbol n\,$  has no physical meaning. In this case, each of the transformations (11)
is differed from  the respective Lorentz boost $\,x'^i=L^i_k(\boldsymbol v) x^k\,$
by the additional rotation
$\,x'^i=R^i_k(\boldsymbol
v,\boldsymbol n ) x^k\,$
of the spatial axes. These additional rotations are adjusted in such a way that if a ray of light has the the direction $\,\boldsymbol n\,$
in one frame, then it will have the same direction in all frames under consideration.
Thus, at $\,b=0\,,$ i.e. within the framework of conventional special relativity, the transformations (11)
represent  an alternative to the Lorentz boosts, however, in contrast to the boosts, they constitute a 3-parameter noncompact group which turns out to be a subgroup of the 6-parameter homogeneous Lorentz group .
Physically such noncompact transformations are realized as follows. First choose as
$\,\boldsymbol n\,$ a direction towards a preselected star and then
perform an arbitrary Lorentz boost by complementing it with such a
turn of the spatial axes that in a new
reference frame the direction towards the star remains unchanged.
The set of the transformations described has just been given by the above-displayed relation (11).

Now let us consider an inhomogeneous group of relativistic symmetry or, in other words, inhomogeneous group of isometries of the flat Finslerian space-time (9).
Since the respective homogeneous noncompact group (10)
is 3-parametric, with inclusion of 1-parameter group of rotations around the preferred direction
 $\,\boldsymbol n\,$   and 4-parameter translation group, the inhomogeneous group of relativistic symmetry of the space-time (9) appears to have eight parameters. To obtain the simplest representation for its generators, it is enough to send the third spatial axis along  $\,\boldsymbol n\,$  and rewrite the homogeneous transformations (10) in the infinitesimal form. As a result, we come to the following eight generators
\begin{equation}\nonumber
\left.
\begin{array}{rcl}
X_1\!\!&\!\!=\!\!&\!\!-(x^1p_0+x^0p_1)-(x^1p_3-x^3p_1)\,,\\
X_2\!\!&\!\!=\!\!&\!\!-(x^2p_0+x^0p_2)+(x^3p_2-x^2p_3)\,,\\
X_3\!\!&\!\!=\!\!&\!\!-bx^ip_i-(x^3p_0+x^0p_3)\,,\\
R_3\!\!&\!\!=\!\!&\!\!x^2p_1-x^1p_2\,; \qquad \qquad \qquad \qquad \qquad p_i=\partial /\partial
x^i\,.\\
\end{array}
\right.
\end{equation}
These generators satisfy the commutation relations
\begin{equation}\nonumber
\left.
\begin{array}{llll}
  [X_1X_2]=0\,, & [R_3X_3]=0\,, && \\
  \left[X_3X_1\right]=X_1 \,, & [R_3X_1]=X_2 \,, && \\
  \left[X_3X_2\right]=X_2\,, & [R_3X_2]=-X_1\,; & &   \vspace*{10pt}\\
  \vspace*{10pt}
  \left[p_i p_j\right]=0\,; &&& \\
  \left[X_1p_0\right]=p_1\,,& [X_2p_0]=p_2\,, & [X_3p_0]=bp_0+p_3\,, &
  [R_3p_0]=0\,, \\
  \left[X_1p_1\right]=p_0+p_3\,,& [X_2p_1]=0\,, & [X_3p_1]=bp_1\,, &
  [R_3p_1]=p_2\,, \\
  \left[X_1p_2\right]=0\,, & [X_2p_2]=p_0+p_3\,, & [X_3p_2]=bp_2\,, &
  [R_3p_2]=-p_1\,, \\
  \left[X_1p_3\right]=-p_1\,, & [X_2p_3]=-p_2\,, & [X_3p_3]=bp_3+p_0\,, &
  [R_3p_3]=0\,. \\
\end{array}
\right.
\end{equation}
It is clear that the homogeneous isometry group of the flat Finslerian space-time (9)  contains four parameters (\,generators $\,X_1\,,\, X_2\,,\, X_3\,$
and $\,R_3\,$). This group is a subgroup of the 11-parametric Similitude (Weyl) group [8], and it is isomorphic to the corresponding 4-parametric subgroup (\,with generators $\,X_1\,,\, X_2\,,\, X_3\!\!\mid_{b=0}\,$  and  $\,R_3\,$) of the homogeneous Lorentz group. Since the 6-parametric homogeneous Lorentz group does not have any 5-parametric subgroup [9], while its 4-parametric subgroup is unique up to isomorphisms, the passage from the Minkowski space-time to the Finslerian space-time (9) implies a minimum possible violation of Lorentz symmetry. With this, the relativistic symmetry represented now by generalized Lorentz boosts (10)
remains valid [10]. Besides, in spite of a new geometry of event space, the Lobachevsky geometry of 3-velocity space and, respectively, the relativistic law of addition of 3-velocities remain unchanged.
Finally, it should be noted that the above-described 8-parameter inhomogeneous group of relativistic symmetry (the group of motions) of the anisotropic event space (9) was first scrutinized in [11]-[13]. At the present time, thanks to [14], this group is known as $\,DISIM{_{b}}(2)\,,$ i.e. Deformed Inhomogeneous  subgroup of SIMilitude group that includes 2-parameter Abelian homogeneous noncompact subgroup.\footnote{In connection with the $\,DISIM{_{b}}(2)$ relativistic symmetry, note here a recent interesting paper [23].} As for the group $\,SIM(2)\,,$ its generators $\,X_1\,,\, X_2\,,\, X_3\!\!\mid_{b=0}\,,\, R_3\,$ and their Lie algebra have already been described above. The Lie algebra of the corresponding Hermitian operators is also displayed in [15], where  a nontrivial physical implementation of the $\,SIM(2)\,$ symmetry is proposed. \\

\noindent
{\bf 3.\, Relativistic point mechanics and the rest momentum as a new observable
%\phantom{in t}
in the flat Finslerian space-time with partially broken 3D isotropy}\\

A remarkable property of the partially anisotropic space of events (9)
is the fact
that it keeps the conformal structure (\,light cone\,) of the Minkowski space,
i.e. light propagates according to the equation $dx_0^2-d\boldsymbol x^{2}=0\,.$
Therefore, the speed of light is independent of the direction of its
propagation and is equal to $c\,$. It thus appears that the square of the
distance $\,dl\,$ between adjacent points of 3D space, determined by
means of exchange of light signals,
is expressed by the formula $dl^2=d\boldsymbol x^{2}\,$. Thus, although in
the 3D space there is a preferred direction $\,\boldsymbol n\,,$ its geometry
remains Euclidean. But, what does the anisotropy of space physically manifest
itself in?
First of all, the space anisotropy affects the dependence of proper time of a moving clock on the direction of clock's velocity\,: according to (9), the interval $\,d\tau\,$ of
proper time read by the clock moving with a velocity $\boldsymbol v\,$
is related to the time interval $\,dt\,$ read by a clock at rest by the relation
$\,d\tau =(d\tau /dt)\,dt\,,$ where
$$
\frac{d\tau}{dt} =\left(\frac{1-\boldsymbol v\boldsymbol n /c}{
\sqrt{1-\boldsymbol v^{\,2}/c^2}}\right)^{\!\!b}\sqrt{1-\boldsymbol v^{\,2}/c^2}\,\,.
$$
Thus, in contrast to Minkowski space (\,for which:
 $b=0\,,$ $(d\tau /dt)\!\!\mid_{b=0}\,=\sqrt{1-v^2/c^2}\le 1\,$  and, hence,
the moving clock is
always slow in comparison with the clock at rest\,), in the anisotropic space the
time dilatation factor $\,(d\tau /dt)\!\!\mid_{b>0}\,$  can take on values
greater than unity.\\
\begin{figure}[hbt]
\vspace*{-0.2cm}
%\hspace*{-1.2cm}
\begin{center}
\epsfig{figure=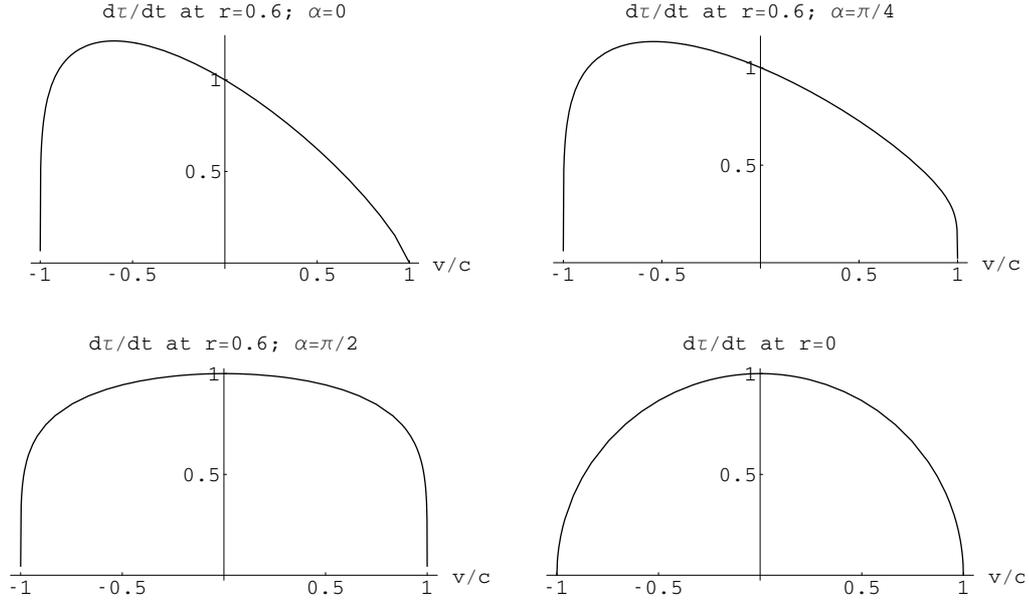,scale=0.8}
\end{center}
\caption{Plots illustrating the dependence of time dilatation factor $\,(d\tau /dt)\!\!\mid_{b\equiv r=0.6}\,$ on the magnitude of clock's velocity  at  angles $\,\alpha\,$ between $\,\boldsymbol v\,$ and $\,\boldsymbol n\,$ equal to $\,0\,,\, {\pi}/4\,$ and $\,{\pi}/2\,.$ The semi-circumference corresponds to  the isotropic Minkowski space for which $b\equiv r=0$.}
\end{figure}
In order to generalize conventional relativistic point mechanics for the case of the flat partially anisotropic
space-time, we replace the Minkowski line element $ds\!=\!\sqrt{dx_0^2-d\boldsymbol x^2}$
with the Finslerian element in the usual action integral.
As a result, the action integral for a free relativistic particle in the flat partially anisotropic space-time takes, according to (9), the form
\begin{equation}\label{12}
S=-mc\int\limits_i^fds\,,
\end{equation}
where
$
ds=\left[(dx_0-\boldsymbol n d\boldsymbol
x)/{\sqrt{dx_0^2-d\boldsymbol x^2}}\right]^{b} \sqrt{dx_0^2-d\boldsymbol x^2}\,.
$
Now we have to make sure that formula (12) correctly determines the action for a free particle.
To this end, consider, along with the clock at rest, another clock which starts to move from the clock at rest and, having executed nonuniform motion along a closed path, returns to the initial point. Since at certain values of the velocity $\,\boldsymbol v\,$  the time dilatation factor in the anisotropic space is greater than unity (\,see Figure 1\,), the moving clock is fast (on the corresponding parts of its path) in comparison with the clock at rest.  Yet, on returning to the clock at rest, the moving clock will necessarily be behind. The validity of this assertion was rigorously proved in [16]. Thus in the flat anisotropic space-time (9) the integral  $\,\int\limits_i^fds\,$ is at a maximum while the action (12) is at a minimum on the straight world line connecting  points $\,i\,$ and $\,f\,.$ Therefore, (12) really correctly determines the action for a free particle in the flat anisotropic space-time (9).

The action (12) leads to the following relativistic Lagrange function
\begin{equation}\label{13}
L = -mc^2\left (\frac{1-\boldsymbol v\boldsymbol n /c}{\sqrt{1-\boldsymbol v^{\,2}/c^2}
}\right )^{\!\!b}
\,\sqrt{1-\boldsymbol v^{\,2}/c^2}\,.
\end{equation}
It follows from (13) that
\begin{equation}\label{14}
E = \frac{mc^2}{\sqrt{1\!-\!\boldsymbol v^{2}\!/\!c^2}}\left (\frac{1\!-\!\boldsymbol
v\boldsymbol n \!/\!c}{
\sqrt{1\!-\!\boldsymbol v^{2}\!/\!c^2}}\right )^{\!\!b}\left [1\!-\!b\!+\!b\,\frac{1\!-\!\boldsymbol v^{2}\!/\!c^2}{
1\!-\!\boldsymbol v\boldsymbol n \!/\!c}\right ]\,.
\end{equation}
Thus, the energy $\,E\,$ of a free particle in the anisotropic space depends both on the magnitude and on the direction of its velocity $\,\boldsymbol v\,.$  At $\boldsymbol v=0\,$ the energy reaches its absolute minimum, i.e the rest energy $\,E\!\!\mid_{\boldsymbol v=0}\,=mc^2\,$. As regards the momentum
\begin{equation}\label{15}
\boldsymbol p = \frac{m}{\sqrt{1\!-\!\boldsymbol v^{2}\!/\!c^2}}\left (\frac{1\!-\!\boldsymbol v\boldsymbol n \!/\!c}{
\sqrt{1\!-\!\boldsymbol v^{2}\!/\!c^2}}\right )^{\!\!b}\left [(1\!-\!b)\boldsymbol v\!+\!bc\boldsymbol n\frac{1\!-\!\boldsymbol v^{2}\!/\!c^2}{
1\!-\!\boldsymbol v\boldsymbol n \!/\!c}\right ]\,,
\end{equation}
its direction  does not
coincide with the direction of the velocity $\,\boldsymbol v\,$ of a massive particle. Even
in the case $\,\boldsymbol v=0\,$ the momentum of the particle does not vanish; there
remains its rest momentum. Thus, in the anisotropic space (9), any massive particle besides its rest energy $E\!\!\mid_{\boldsymbol v=0}\,=mc^2$ also has the rest momentum ${\boldsymbol p}\!\!\mid_{\boldsymbol v=0}\,=bmc\boldsymbol n .$ Massless particles
have no such property; for them, as in conventional special relativity,
$\,v\!=\!c\,$ and $\,E^2\!/\!c^2\!-\!{\boldsymbol p}^{\,2}\!=\!0.$
%\newpage$\,(d\tau /dt)\!\!\mid_{b\equiv r=0.6}\,$

The energy (14) and the momentum (15) are
related by the following modified dispersion relation (\,modified mass-shell equation\,)
\begin{equation}\label{16}
\left[ {\frac{(p_0-{\boldsymbol {pn}})^2}{p_0^2-{\boldsymbol p}^2}}\right]^{-b}
(p_0^2-{\boldsymbol
 p}^2) = {(mc)}^2(1-b)^{(1-b)} (1+b)^{(1+b)}\,,
\end{equation}
where $\,p_0=p^0=E/c\,.$
This modified relation describes a deformed two-sheeted hyperboloid. Such a hyperboloid is relativistically invariant owing to the fact that the generalized Lorentz transformations (10) of the event coordinates $\,x^i\,$, i.e. the transformations
$
\,x'^i=D R^i_j L^j_k x^k\,
$
are accompanied by the corresponding transformations
$
\,p'^i=D^{-1} R^i_j L^j_k p^k\,
$
of 4-momenta $\,p^i=(\,p^0, \boldsymbol p\,)\,.$ It is obvious that the deformation of a two-sheeted hyperboloid depends on the magnitude $\,b\,$ of space anisotropy. The corresponding two-sheeted hyperboloids at $\,b\equiv r = 0\,, 0.3\,, 0.6$ and 0.8 are shown in Figure 2.

\begin{figure}[hbt]
%\vspace*{-0.2cm}
%\hspace*{-1.2cm}
\begin{center}
\epsfig{figure=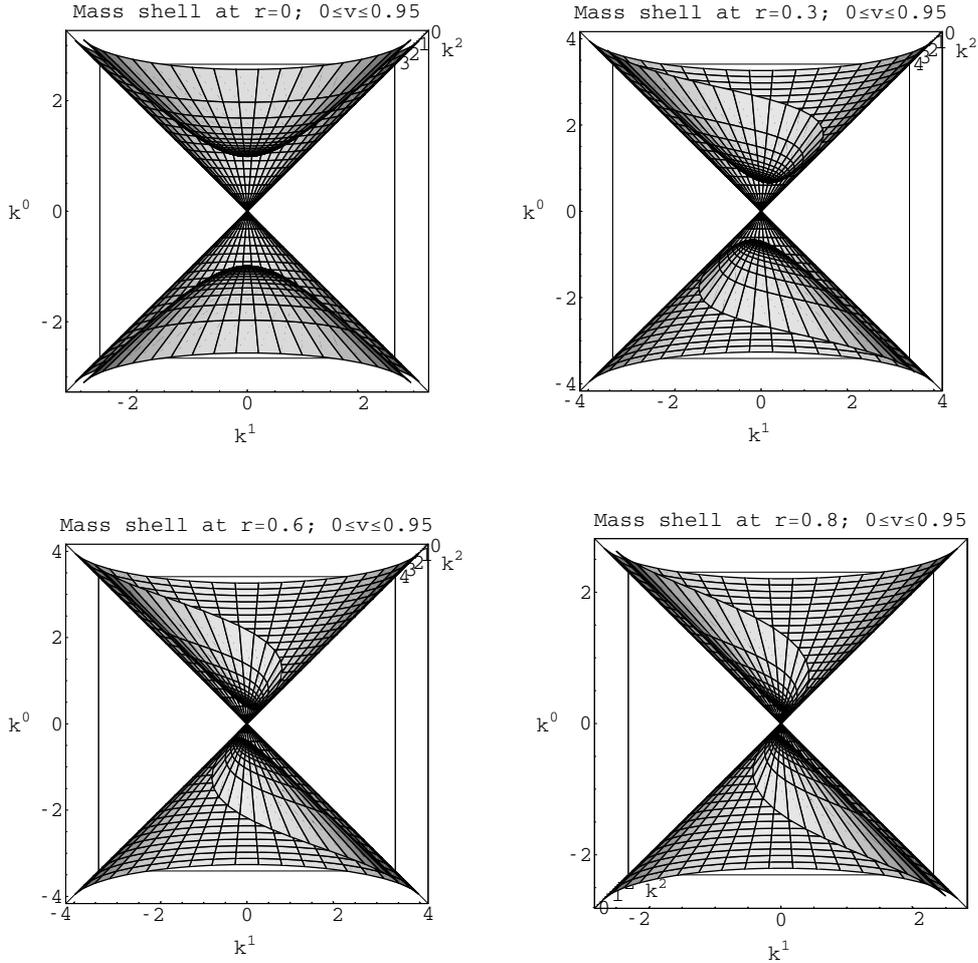,scale=0.8}
\end{center}
\vspace*{-0.6cm}
\caption{Parametric 3D plots of the mass shells in a space of kinematic 4-momenta. Any of the
deformed two-sheeted hyperboloids remains inscribed into a light cone and like a light cone
it is an invariant of the generalized Lorentz transformations.}
\end{figure}
The Lagrange function (13) takes on the following form in the nonrelativistic limit
\begin{equation}\label{17}
L=-mc^2+bmc(\boldsymbol v\boldsymbol n )+(1-b)m{\boldsymbol v}^{\,2}/2+b(1-b)m(\boldsymbol v\boldsymbol n )^2/2\,.
\end{equation}
Since in (17) the expression $\,-mc^2+bmc(\boldsymbol v\boldsymbol n )\,$  is a total derivative with respect to time, it can be omitted. As a result, we see that in the case of a nonrelativistic particle in the anisotropic space its  kinetic energy  $\,T=m_{\alpha\beta}\,v^\alpha v^\beta /2\,$ and momentum $\,p_\alpha =m_{\alpha\beta}\,v^\beta \,$ are determined by the inertial mass tensor
\begin{equation}\label{18}
m_{\alpha\beta}=m(1-b)({\delta}_{\alpha\beta}+b\,{n}_\alpha{n}_\beta )\,.
\end{equation}

Let us now rewrite  Finslerian metric (9) so that it is expressed through the four-dimensional entities:
\begin{equation}\label{19}
ds=\left[\frac{(dx_0-\boldsymbol n d\boldsymbol
x)^2}{dx_0^2-d\boldsymbol x^{2}}\right]^{\!b/2}\!\!\sqrt{dx_0^2-d\boldsymbol
x^2}=\left[\frac{(n_idx^i)^2}{{\eta}_{ik}dx^idx^k}\right]^{\!b/2}\!\!\sqrt{{\eta}_{ik}dx^idx^k}\,.
\end{equation}
Since $\,{\boldsymbol n}^2=1\,,$ it is clear that here we have
\begin{equation}\nonumber
n_i=\{1,-\boldsymbol n\}\,,\quad\!\! {\eta}_{ik}=diag\{1,-1,-1,-1\}\,,\quad\!\! n^i=\{1,\boldsymbol n\}\,,\quad\!\! n_in^i=0.
\end{equation}
In accordance with (12) and (19), a constant field $\,b\,$ defines the specific inseparable
interaction of a constant null-vector field $\,n_i\,$ with massive particles. Effect
of this interaction is that the particles acquire - according to (18), the properties of quasi-particles
in an axially symmetric crystalline medium, the role of which is played by axially symmetric quintessence.
The latter is, thus, the physical carrier of the anisotropy of the flat space of events (19)
and it consists of Weyl fermion-antifermion condensate.
In this regard, it is worth noting that, in contrast to the standard Dirac equation, the corresponding constant spinors are non-trivial solutions of the $\,DISIM{_{b}}(2)-$invariant generalized massive Dirac
equation [17], whose Lagrangian has the form
\begin{equation}\label{20}
{\cal L}=\frac{i}{2}\left(\bar\psi{\gamma}^j{\partial}_j\psi
-{\partial}_j \bar\psi{\gamma}^j\psi\right) -
m\left[\left( \frac{n_j \bar{\psi} \gamma^j
\psi}{\bar{\psi}\psi} \right)^{2} \right]^{b/2}\bar{\psi}\psi  \,.
\end{equation}

\noindent
{\bf 4.\, Metric and group of relativistic symmetry of the flat Finslerian space-time
with entirely broken 3D isotropy}\\

In [18] the metric describing relativistically invariant Finslerian spaces with entirely
broken 3D isotropy has been obtained in the following most general form
\begin{equation}\label{21}
\begin{array}{rl}
ds=&\!\!\!(dx_0-dx_1-dx_2-dx_3)^{(1+b_1+b_2+b_3)\,/\,4}
(dx_0-dx_1+dx_2+dx_3)^{(1+b_1-b_2-b_3)\,/\,4}\\
\times &\!\!\!(dx_0+dx_1-dx_2+dx_3)^{(1-b_1+b_2-b_3)\,/\,4}
(dx_0+dx_1+dx_2-dx_3)^{(1-b_1-b_2+b_3)\,/\,4}\,.
\end{array}
\end{equation}
The parameters characterizing the anisotropy of the spaces (21) are limited by the conditions
$$
\begin{array}{ll}
1+b_1+b_2+b_3\ge 0\,,&1+b_1-b_2-b_3\ge 0\,,\\
1-b_1+b_2-b_3\ge 0\,,&1-b_1-b_2+b_3\ge 0
\end{array}
$$
It should be noted that if $\,b_1=b_2=b_3=0\,$, then the metric (21) takes the form
\begin{equation}\nonumber
\begin{array}{rl}
ds_{_{B-M}}=&[\,\,\,(dx_0-dx_1-dx_2-dx_3)(dx_0-dx_1+dx_2+dx_3)\\
&\times (dx_0+dx_1-dx_2+dx_3)(dx_0+dx_1+dx_2-dx_3)\,\,\,]^{\,1/4}\,.\\
\end{array}
\end{equation}
Thus, in this particular case, we obtain the well-known Berwald-Moor metric, but written
in the basis, which was introduced in [18].

Now consider the group of isometries of flat Finslerian spaces (21). The homogeneous
3-parametric non-compact group of isometries, i.e. the group of the relativistic symmetry
of space-time (21) appears to be Abelian, and the transformations belonging to such a
group have the same meaning as the ordinary Lorentz boosts. The explicit form of these
transformations is
\begin{equation}\label{22}
x'_i=D\,L_{ik}\,x_k\,,
\end{equation}
where
\begin{equation}\nonumber
D=e^{-(\,b_1\,\alpha _1+b_2\,\alpha _2+b_3\,\alpha _3\,)}\ ,
\end{equation}
$L_{ik}$ are the unimodular matrices that are given by the formulas
\begin{equation}\label{23}
L_{ik}=\left (
\begin{array}{rrrr}
\cal A&-\cal B&-\cal C&-\cal D\\
-\cal B&\cal A&\cal D&\cal C\\
-\cal C&\cal D&\cal A&\cal B\\
-\cal D&\cal C&\cal B&\cal A\\
\end{array}
\right )\ ,
\end{equation}	
$$
{\cal A}=\cosh \alpha _1\cosh \alpha _2\cosh \alpha _3+
\sinh \alpha _1\sinh \alpha _2\sinh \alpha _3\,,
$$
$$
{\cal B}=\cosh \alpha _1\sinh \alpha _2\sinh \alpha _3+
\sinh \alpha _1\cosh \alpha _2\cosh \alpha _3\,,
$$
$$
{\cal C}=\cosh \alpha _1\sinh \alpha _2\cosh \alpha _3+
\sinh \alpha _1\cosh \alpha _2\sinh \alpha _3\,,
$$
$$
{\cal D}=\cosh \alpha _1\cosh \alpha _2\sinh \alpha _3+
\sinh \alpha _1\sinh \alpha _2\cosh \alpha _3\,\phantom{,}
$$
$\alpha _1\,,\alpha _2\,,\alpha _3\,$ are the parameters of the group. Along with the parameters $\alpha_i\,,$  the components
$\,v_i = dx_i/dx_0\,$ of the coordinate velocity of the primed reference frame can also be used as
group parameters. The parameters $\,v_i\,$  and $\alpha_i\,$ are related by
\begin{equation}\label{24}
\begin{array}{rrr}
& &v_1=(\tanh\alpha _1-\tanh\alpha_2\tanh\alpha _3)/(
1-\tanh\alpha_1\tanh\alpha_2\tanh\alpha_3)\,,\\
& &v_2=(\tanh\alpha _2-\tanh\alpha_1\tanh\alpha _3)/(
1-\tanh\alpha_1\tanh\alpha_2\tanh\alpha_3)\,,\\
& &v_3=(\tanh\alpha _3-\tanh\alpha_1\tanh\alpha _2)/(
1-\tanh\alpha_1\tanh\alpha_2\tanh\alpha_3)\,.
\end{array}
\end{equation}
The reverse relations have the form
\begin{eqnarray*}
& &\alpha _1\,=\frac{1}{4}\ln \frac{(1+v_1-v_2+v_3)(1+v_1+v_2-v_3)}
{(1-v_1-v_2-v_3)(1-v_1+v_2+v_3)}\,\,,\\
& &\alpha _2\,=\frac{1}{4}\ln \frac{(1-v_1+v_2+v_3)(1+v_1+v_2-v_3)}
{(1-v_1-v_2-v_3)(1+v_1-v_2+v_3)}\,\,,\\
& &\alpha _3\,=\frac{1}{4}\ln \frac{(1-v_1+v_2+v_3)(1+v_1-v_2+v_3)}
{(1-v_1-v_2-v_3)(1+v_1+v_2-v_3)}\,\,.
\end{eqnarray*}
As for the generators $X_i$ of the homogeneous 3-parametric group of isometries (22) of the
space-time (21), they can be represented as follows
\begin{gather*}\nonumber
 X_1=-b_1x_{\alpha}p_{\alpha}-(x_1p_0+
x_0p_1)+(x_2p_3+ x_3p_2),\\\nonumber
 X_2=-b_2x_{\alpha}p_{\alpha}-(x_2p_0+
x_0p_2)+(x_1p_3+
x_3p_1),\\
X_3=-b_3x_{\alpha}p_{\alpha}-(x_3p_0+
x_0p_3)+(x_1p_2+ x_2p_1),
\end{gather*}
where $p_{\alpha}=\partial /\partial x_{\alpha}$ are the generators of the 4-parametric group of translations. Thus,
with inclusion of the latter, a inhomogeneous group of isometries of the entirely anisotropic
Finsler space of events (21) is a 7-parametric group. As to its generators, they satisfy the
commutation relation
\begin{alignat*}{4}
 & \left[X_i X_j\right]=0,\qquad && \left[p_{\alpha} p_{\beta}\right]=0, && & \nonumber\\
 & \left[X_1p_0\right]=b_1p_0+p_1,\qquad && [X_2p_0]=b_2p_0+p_2,\qquad  && [X_3p_0]=b_3p_0+p_3, &\nonumber\\
 & \left[X_1p_1\right]=b_1p_1+p_0,\qquad && [X_2p_1]=b_2p_1-p_3,\qquad && [X_3p_1]=b_3p_1-p_2, & \nonumber\\
 & \left[X_1p_2\right]=b_1p_2-p_3,\qquad & & [X_2p_2]=b_2p_2+p_0, \qquad && [X_3p_2]=b_3p_2-p_1, &\nonumber\\
 & \left[X_1p_3\right]=b_1p_3-p_2, \qquad && [X_2p_3]=b_2p_3-p_1, \qquad && [X_3p_3]=b_3p_3+p_0. & %\label{eq18}
\end{alignat*}

Determining the distance $\,dl\,$ between adjacent points  by means of light signals leads to the conclusion that the flat entirely anisotropic 3D space is endowed with the discrete symmetry of a regular rhombic dodecahedron shown in Figure 3. Accordingly, the dynamic properties of a nonrelativistic massive particle residing in such an anisotropic space are determined by the inertial mass tensor
\begin{equation}\label{22}
m_{\alpha\beta} =
m\left( \begin{array}{ccc}
(1-{b_1}^2) & (b_3-b_1b_2) & (b_2-b_1b_3) \\
(b_3-b_1b_2) & (1-{b_2}^2) & (b_1-b_2b_3) \\
(b_2-b_1b_3) & (b_1-b_2b_3) & (1-{b_3}^2)
\end{array} \right)\,.
\end{equation}

\vspace*{1cm}
\begin{figure}[hbt]
%\vspace*{-0.2cm}
%\hspace*{2.2cm}
\begin{center}
\epsfig{figure=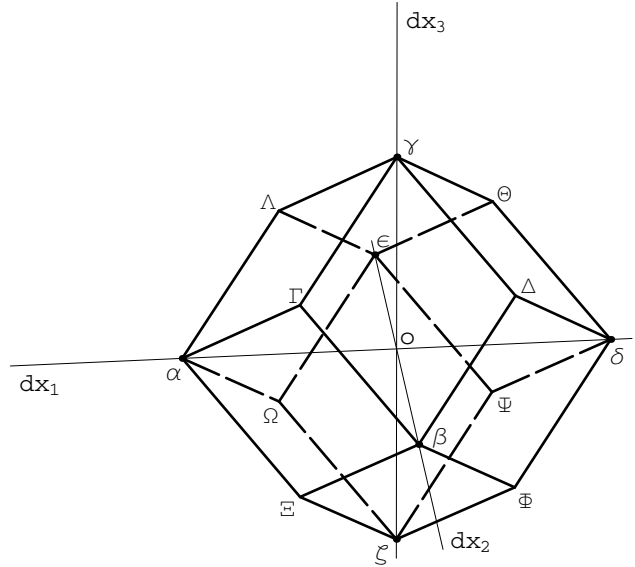,scale=0.8}
\end{center}
\caption{A regular rhombic dodecahedron as an Euclidean image of 3D sphere of
radius $dl$, prescribed in the flat entirely anisotropic Finslerian space.}
\end{figure}

\noindent
{\bf 5.\, A curved space-time with $\,DISIM{_{b}}(2)\,$ local relativistic symmetry and local gauge invariance
                               of its Finslerian metric}\\

Currently, the problem of Lorentz symmetry violation is widely debated in the literature, in which case the Finslerian approach to the problem is becoming more and more popular. It is based on the assumption of the Finsler geometric structure of space-time.
As for Finslerian generalization of the Einstein equations, the corresponding first attempts were made in the early seventies. At that time, equations for the metric function of Finslerian space-time were proposed by analogy with the Einstein equations, but using the Finsler curvature tensors. Unfortunately, the physical interpretation of such equations turned out to be obscure.  Although recent studies are more advanced from a physical point of view ( see, in particular, an interesting paper [24] ), it should again be noted that, in all likelihood, it is impossible to build physically meaningful Finslerian generalization of the Einstein theory of relativity without going beyond the framework of the purely geometric Finslerian analogies.

The fact is that the transition from Riemannian to Finslerian space-time necessarily leads to a violation of the local Lorentz symmetry. However, the violation of the local Lorentz symmetry of space-time does not necessarily mean a violation of the local relativistic symmetry. In the previous sections of this paper, it is shown that there are only two types of flat Finslerian event spaces, transitions to which from the Minkowski space violate the Lorentz symmetry of the event space without breaking relativistic symmetry.\footnote{Since the Lorentz symmetry is usually identified with relativistic symmetry (although, strictly speaking, these symmetries do not coincide), we clarify that relativistic symmetry of a flat event space means that it has at least a 3-parameter homogeneous non-compact group of isometries, while the Lorentz symmetry means that an homogeneous group of isometries is represented by the 6-parameter Lorentz group, which, in addition to 3-parameter non-compact  transformations,
contains the 3-parameter subgroup of 3D rotations.}

The flat Finslerian event spaces of the first type are relativistically symmetric spaces with partially broken 3D isotropy, i.e. with axial 3D symmetry. Their $\,DISIM{_{b}}(2)\,$ invariant Finslerian metric has the form (9) and describes axially symmetric crystalline states of space-time. In addition, such a metric satisfies the correspondence principle with the pseudo-Euclidean metric of the Minkowski space and therefore underlies viable Finslerian extension of the SR.

The flat Finslerian event spaces of the second type are relativistically symmetric spaces with entirely broken 3D isotropy. They are described by relativistically invariant Finslerian metric (21) and their relativistic symmetry is realized by means of 3-parameter homogeneous non-compact Abelian group of isometries. As for the respective non-homogeneous group, it is a 7-parameter group.
It is also worth noting that flat Finslerian metric (21) describes an entirely anisotropic crystalline state of space-time with a discrete symmetry of a regular rhombic dodecahedron.

It is now clear that  Finslerian metric describing a curved locally anisotropic space-time with $\,DISIM{_{b}}(2)\,$  local relativistic symmetry can be obtained from the Finslerian metric (19) of a flat space-time by means of the following replacement\,:
$${\eta}_{ik}\to g_{ik}(x)\,,\,  {n}_i\to {n}_i(x)\,,\, b\to b(x)$$\,.
As a result, we arrive at the only possible viable Finslerian space-time model [19] with the metric
\begin{equation}\label{26}
ds=\left[\frac{(n_idx^i)^2}{g_{ik}dx^idx^k}\right]^{\!b/2}\!\!\sqrt{g_{ik}dx^idx^k}\,,
\end{equation}
where $\,g_{ik}\!=\!g_{ik}(x)\,$ is the Riemannian metric tensor related to the gravity field; $\,b\!=\!b(x)\,$  is the scalar field characterizing the magnitude of local space anisotropy, and $\,n_i\!=\!n_i(x)\,$ is the null-vector field indicating the locally preferred directions in space-time. So, the correct Finslerian extension of the general theory of relativity ( under which the physical meaning of local relativistic space-time symmetry is preserved) must be based on the Finslerian space-time model (26). At each point of a curved Finslerian space time (26), the flat tangent Finslerian spaces have their own values of the parameters $\,b\,$ and $\,{n}_i\,$. These values are none other than the value of the fields $\,b(x)\,$ and $\,{n}_i(x)\,$ at the corresponding space-time points.

Obviously, the dynamics of a curved Finslerian spacetime (26) is completely determined  by the dynamics of interacting fields  $\,g_{ik}(x), b(x), {n}_i(x)\,,$ and these fields form, along with the matter fields, a unified dynamic system. Therefore, in contrast to the existing purely geometric approaches to a Finslerian generalization of the Einstein equations, our approach to the same problem is based on the use of the methods of conventional field theory.

One can readily check by direct substitution that the metric (26), depending on the fields $\,g_{ik}(x)\,$, $\,b(x)\,$, $\,n_i(x)\,$ and thereby  determining a curved partially anisotropic Finslerian space-time, is invariant under the following local  transformations of these fields
\begin{equation}\label{27}
\left\{
\begin{array}{lcl}
g_{ik}(x)\!&\!\rightarrow&{\tilde g}_{ik}(x)=\exp\{2\sigma(x)\}\,g_{ik}(x)\,,\\
n_i(x)\!&\!\rightarrow&{\tilde n}_i(x)=\exp\{\sigma(x)[b(x)\!-\!1]/b(x)\}\,n_i(x)\,,\\
b(x)\!&\!\rightarrow&{\tilde b}(x)=b(x)\,,\\
\end{array}
\right.
\end{equation}
where $\,\sigma(x)\,$   is an arbitrary function.

From the invariance of  metric (26) under  the local  transformations (27) it follows that the Lagrangian and the action
\begin{equation}\label{28}
S=-mc\int\limits_a^b\left[\frac{(n_i(x){\dot x}^i)^2}{g_{ik}(x){\dot x}^i{\dot x}^k}\right]^{\!b(x)/2}\!\!\sqrt{g_{ik}(x){\dot x}^i{\dot x}^k}d\lambda
\end{equation}
(\,here $\,{\dot x}^i\,$ is the generalized velocity, $\,\lambda\,$  is a parameter on a world line\,) for a particle  in the external fields $\,g_{ik}(x)\,$, $\,b(x)\,$ and $\,n_i(x)\,,$  are  also invariant under the above transformations, which entails the invariance of Hamilton's equations of motion
\begin{eqnarray}\label{29}
& &\frac{du_i}{ds}=\frac{1}{2}\left[\frac{\partial b}{\partial x^i}\ln\frac{(1\!+\!b)(n^ku_k)^2}{(1\!-\!b)g^{lm}u_lu_m}+\frac{\partial n^p}{\partial x^i}\frac{2bu_p}{n^ku_k}-\frac{\partial g^{nl}}{\partial x^i}\frac{(1\!+\!b)u_nu_l}{g^{jk}u_ju_k}\right]\,,
\\
& &\frac{dx^i}{ds}=\frac{(1\!+\!b)g^{ik}u_k}{g^{lm}u_lu_m}-b\frac{n^i}{n^ku_k}\,,\label{30}
\end{eqnarray}
where
\begin{equation}\nonumber
u_l=(1\!-\!b)\frac{v_l}{v_kv^k}+b\frac{n_l}{n_kv^k}\,; \qquad v^k= dx^k/ds\,.
\end{equation}
Moreover, it can readily be seen that the observable quantities of  proper time
\begin{equation}\label{31}
cd\tau =\left[\frac{n_{\!_0}^2}{g_{_{00}}}\right]^{b/2}\sqrt{g_{_{00}}}dx^{_0}
\end{equation}
and 3-space distances
\begin{equation}\label{32}
dl^2=\left[\frac{n_{\!_0}^2}{g_{_{00}}}\right]^{b}\left(-g_{_{\alpha\beta}}+\frac{g_{_{0\alpha}}g_{_{0\beta}}}{g_{_{00}}}\right)dx^{\alpha}dx^{\beta}\,,
\end{equation}
as well as the action
\vspace*{-2mm}
$$
S=-\frac{1}{c}\int\! \mu ^\ast\!
\left(\frac{n_i\,v^i}{\sqrt{\,g_{ik}\,v^i\,v^k}}\right )^{\!4b}\!\sqrt{-g}
\,\,d^{\,4}x\,,
$$
for a  compressible fluid are  invariant under (27). ( In this formula $\mu ^\ast$ is  the invariant fluid energy  density,  $v^i=dx^i/ds$, and $ds$
 is Finslerian metric (26) )\\
All this suggests that the fields coupled by the transformations (27) describe the same physical situation and that the transformations (27) themselves have the meaning of local gauge transformations.

In connection with the mentioned local gauge invariance, the dynamical system
consisting of the fields $g_{ik}, b , n^i$ and a compressible fluid must be supplemented by two
vector gauge fields $A_i$ and $B_i$ , that under local transformations (27) are transformed in
the corresponding gradient manner. The $A_i$ field for a certain class of problems is a pure
gauge field, and the $B_i$ field, whose gauge transformation has the form
$$
B_i \rightarrow B_i + l[(b-1)\sigma (x)/b]_{;i}\,,
$$
where $l$ is a constant with the dimensionality of length, interacts with the conserved
rest mass current $j^i$ , adding the term proportional to $B_ij^i$ to the full gauge invariant
Lagrangian.\\

\noindent
{\bf 6.\, Conclusion}\\

Alongside with the phenomena, for explanation of which the concept
of dark matter or dark energy is commonly used, there are the data
of astrophysical observations and particle physics, indirectly indicating
the existence of a local space-time anisotropy in the Universe. In this
respect, we should mention first of all the anisotropy of the accelerated expansion of the Universe and the possible anisotropy of the fine-structure constant,
a unified Finslerian description of which was proposed in [20]. Note also the
Greisen-Zatsepin-Kuzmin effect in the physics of ultra-high energy cosmic rays and
other effects considered in [21]. In particular, the absence of the GZK cutoff has
yet not been explained convincingly and still remains the main empirical fact which
indirectly speaks in favor of violation of Lorentz symmetry and, therefore, in favor of
local space-time anisotropy.

Actually, the first  direct evidence  of the existence of local space-time anisotropy was obtained by the CMS collaboration at LHC within the framework of the so-called Ridge/CMS effect [22]. Let us note in passing that the Ridge/CMS effect is characteristic only to the high multiplicity events. Such events take place in case of the central collision of the initial protons where the energy density at the moment of the collision is comparable to the energy density shortly after the Big Bang, when instead of hadrons there was quark-gluon plasma. Since the initial
total momentum within the CMS experiment is equal to zero, the appearance of a preferred direction, which coincides with  the direction of an elongated ridge of measured correlation function and with the protons collision axis, speaks of vacuum rearrangement with the appearance of
anisotropic axially-symmetric condensate. On the
one hand, quantum-field vacuum filled with the anisotropic condensate
is the physical carrier of the local anisotropy of space-time and it can be
regarded as an anisotropic quintessence, on the other -- it imparts all the
particles the properties of quasi-particles in the crystalline environment.
In particular, in addition to the rest energies, all massive particles
acquire the respective rest momenta $\,{\boldsymbol p}\!\!\mid_{\boldsymbol v=0}\,=bmc\boldsymbol n \,$ and the anisotropy (18) of their inertial properties.

Generally speaking, in contrast to the Minkowski space with its constant quintessence, the flat Finslerian space (9) and its anisotropic quintessence
are able to change their anisotropy if  $\,b\,$  is changeable.
It easy to see that, at $\,b\to 1\,,$  the Finslerian metric $
ds=\left[(dx_0-\boldsymbol n d\boldsymbol
x)/{\sqrt{dx_0^2-d\boldsymbol x^2}}\right]^{b} \sqrt{dx_0^2-d\boldsymbol x^2}\,
$
degenerates into the total differential $\,ds=dx_0-\boldsymbol n d\boldsymbol x\,.$
This means that the notion of spatial extent disappears and in the space-time there remains the
single physical characteristic, namely, time duration and it should be
regarded as an interval of absolute time. Besides at $\,b\to 1\,$
in accordance with (18) there also disappears the inertial mass of any particle.
All this suggests that absolute time, where the very notions
of spatial extent and inertial mass together with unobservable primordial quintessence become meaningless, is not a stable degenerate state of space-time.
As a result of geometric phase
transition, which accompanies a spontaneous breaking of the original gauge symmetry,
such a primordial space-time may turn  into anisotropic space-time, the anisotropy of which and, accordingly, the anisotropy of the quintessence decrease in the course of  the accelerated expansion of the Universe.

In this regard, it is important to bear in mind that due to the directly proportional dependence of the rest momentum on the magnitude of local space-time anisotropy, as well as on mass of a body and speed of light, the fact that in our time the magnitude of the local space-time anisotropy is extremely small does not at all mean the impossibility of detecting anisotropy in astrophysical processes with a giant energy release and, accordingly, with a giant mass defect. Obviously, in the processes with giant mass defects, one can also expect noticeable momentum defects. With allowance for the local law of conservation of momentum, this implies a possible self-acceleration of matter and local isotropization of the Universe.
If this scenario proves to be correct, then the question
posed by the authors of work [14], namely, why are the magnitude of the local space
anisotropy $\,b\,$ and the $\,\Lambda$-term  simultaneously so small, may be answered.\\

\vspace*{-0.2cm}
\begin{small}
\renewcommand{\refname}

\end{small}

\end{document}